\documentclass[11pt,openany,a4paper]{article}
\usepackage[T2A]{fontenc}
\usepackage[english]{babel}% включение переносов

\usepackage{epigraph}
\usepackage{indentfirst}
\usepackage[unicode=true]{hyperref}
\usepackage{xr}
\usepackage{makeidx}
\makeindex
\usepackage{graphicx}
\usepackage[font=small,labelfont=bf]{caption}

\usepackage{picins}

\usepackage{latexsym}
\usepackage{amsfonts}
\usepackage{amsmath}
\usepackage{amssymb}

\font\twmsbm=msbm10 scaled 1200 \font\nmsbm=msbm9
\newfam\bbold
\textfont\bbold=\twmsbm \scriptfont\bbold=\nmsbm
\scriptscriptfont\bbold=\nmsbm

\font\twscr=rsfs10 scaled 1200 \font\nscr=rsfs10
\newfam\script
\textfont\script=\twscr \scriptfont\script=\nscr
\scriptscriptfont\script=\nscr

\newcommand{\be}{\begin{equation}}
\newcommand{\ee}{\end{equation}}
\newcommand{\bean}{\begin{eqnarray*}}
\newcommand{\eean}{\end{eqnarray*}}

\newcommand{\bea}{\begin{eqnarray}}
\newcommand{\eea}{\end{eqnarray}}

\newcommand{\Div}{{\rm div\,}}

\newcommand{\rot}{{\rm rot\,}}

\newcommand{\p}{\partial}

\renewcommand{\thesection}{\arabic{section}.}

\author{M. G. Ivanov\thanks{Mikhail G. Ivanov. e-mail: {\tt \href{email://ivanov.mg@mipt.ru}{ivanov.mg@mipt.ru}}}\\
\small Moscow Institute of Physics and Technology\\
\small Dolgoprudny, Russia
}
\title{PHYSICS AND TECHNOLOGY SYSTEM OF UNITS FOR ELECTRODYNAMICS\footnote{The paper is extended version of the paper 
\cite{mgi-ef}, published in Russian.}}
\date{}
\begin{document}
\maketitle

\abstract{The contemporary practice is to favor the use of the SI units for electric circuits and the
Gaussian CGS system for electromagnetic field. A modification of the Gaussian system of units 
(the Physics and Technology System) is suggested.
In the Physics and Technology System the units of measurement for electrical
circuits coincide with SI units, and the equations for the electromagnetic field are almost
the same form as in the Gaussian system.  The XXIV CGMP (2011) Resolution <<On the possible future revision of the International System of Units, the SI>> provides
a chance to initiate gradual introduction of the Physics and Technology System as a new
modification of the SI.}

%\tableofcontents

\section{Introduction. One and a half century dispute}

The problem of choice of units for electrodynamics
dates back to the time of M.~Faraday (1822--1831) and J.~Maxwell (1861--1873).
Electrodynamics acquired its final form only after geometrization
of special relativity by H.~Minkowski (1907--1909).
The improvement of contemporary (4-dimentional relativistic covariant)
formulation of electrodynamics and its implementation in practice
of higher education stretched not less than a half of century.
Overview of \emph{some} systems of units, which are used in electrodynamics
could be found, for example, in books \cite{gladun,trunov2011} and paper \cite{babel}.

Legislation and standards of many countries recommend to use in science
and education The International System of Units (SI).
Electrodynamical units of SI originates from the practical system of units, which
was established at The First International Congress of Electricians (Paris, 1881).
Units of practical system are multiples of CGSM units 
(the CGSM is one of old and now obsolete version of the CGS system of units),
which were too small for practical applications.

Nevertheless, the using of the SI in electrodynamics is still a matter of objections.
The Gaussian (symmetrical) CGS-system (CGS) is very popular among physicists.
The CGS is more consistent with symmetries of electrodynamics.
The CGS is a standard for scientific publications and textbooks on theoretical physics. 

One could (with some exaggerating) declare that the SI is system for measurements and the CGS is system 
for formulae and analytical calculations.

Today the standard for electrical circuits is the SI, but for electromagnetic field the standard is the CGS.
E.g. the CGS is used in classical textbook by I.Ye.~Tamm ``Basic theory of electricity'' \cite{tamm},
but the practical units (the predecessors of electrical units of the SI) are used for the alternating current (\S~80 of \cite{tamm}).
The similar preferences are common not only for the classics of science, but for today students and professors
in Moscow Institute of Physics and Technology.

The problem of choice of units for electrodynamics is still actual, moreover, it becomes more actual.
The International Bureau of Weights and Measures in the last 8-th edition of its official
brochure ``The International System of Units (SI)'' \cite{si8} admits
(``Units outside the SI'', the 2-nd paragraph, some word are highlighted by italic font by M.I.)
\begin{quote}
\emph{Individual} scientists should also have the freedom to \emph{sometimes}
use non-SI units for which they see a \emph{particular} scientific advantage in \emph{their} work. An
example of this is the use of CGS-Gaussian units in electromagnetic theory applied to
quantum electrodynamics and relativity.
\end{quote}

Reading this text one has to remember, that coherent modern representation
of the electrodynamics without special relativity is impossible.
It is clear, that the main message of this disclaimer applies primarily to
usage of the CGS system.

The previous 7-th edition of the brochure ``The International System of Units (SI)'' \cite{si7}
does not contain this disclaimer.

The softer attitude towards the CGS system is probably associated with the
planned change of definition of kilogram, ampere, mole and kelvin in the SI by
fixing exact numerical values of the Plank constant, the elementary charge,
the Avogadro constant and the Boltzmann constant \cite{resolution2011}.
This redefinition requires new exact measurements and calculations, which use methods of
quantum electrodynamics.
Just in quantum electrodynamics the CGS looks preferable in comparison with the SI
for many physicists.

\section{Criticism of the SI}

Usage of the SI in theoretical research is complicated.
The SI-units of electrodynamic quantities do  
not consistent with symmetries of the theory, which
related with special relativity.
This inconsistency is related with different dimensions
of the electric and magnetic fields $\mathbf{E}$  and $\mathbf{H}$ 
and the inductions $\mathbf{D}$ and $\mathbf{B}$.
It makes difficult usage of the SI in the teaching of electrodynamics,
especially in the cases where a student has to have a solid understanding 
of the theory structure.
Because of this reason theoretical physics courses in leading Russian universities traditionally use the CGS.

The problem was considered in paper by D.V.~Sivukhin ``The international system of physical units'' \cite{sivuxin-ufn} (see also \cite{sivuxin-electro}, where the comparison of the SI and the CGS is presented),
published in 1979 in the journal ``Soviet Physics Uspekhi''
\emph{by the decision of the Division of General Physics and Astronomy of the USSR Academy of Sciences}.
That practically means the unified position of community of Russian physicists.
We present a vivid quote from this article:
\begin{quote}
  In this respect, the SI system is no more logical than, say, a system in which the length, width,
  and height of an object are expressed not only in terms of different units but have different dimensions as well.
\end{quote}

The inconsistence of the SI with the symmetries of electrodynamics
due to historical reasons, because the foundations of the system were
set before the creation of special relativity.
Moreover, the units of volt, ampere, ohm, farad, etc.
(ascending to the practical system) are
intensively used in technics, are involved in the SI 
and are not involved in the CGS\footnote{In the beginning of XX century in Russia capacitors were fabricated, 
labeled in the CGS system. The CGS unit of capacitance is the centimeter \cite{bulygin}.}.
When in 1948 these units were introduced in the SI, role of the special relativity for electrodynamics was still 
not sufficiently understood by many physicists-experimenters and engineers.
You could ever dream of how to change these units in the early XX century,
but now this units and related standards are widely used not only in measuring devices, but in the whole technique, 
including home electronics.
\emph{This makes it virtually impossible any revision of the SI, which excludes from the system ampere, as the basic unit of.}

Can we combine the wishes of the engineers and theorists?
Yes, we can!

\section{What system of units do we need?}

\subsection{The wishes of a theorist}

\begin{itemize}
\item Electric field strength $\mathbf{E}$ and induction $\mathbf{D}$ have to be of the same dimension, and
in the vacuum \mbox{$\mathbf{E}=\mathbf{D}$}.
\item Electric field induction $\mathbf{B}$ and strength $\mathbf{H}$ have to be of the same dimension, and
in the vacuum \mbox{$\mathbf{B}=\mathbf{H}$}.
\item Fields $\mathbf{E}$ and $\mathbf{B}$ have to be of the same dimension, and
in the vacuum for flat traveling wave \mbox{$|\mathbf{E}|=|\mathbf{B}|$}.
\item 
Magnetic field of a moving charge is a relativistic effect,
so the formula \emph{have to} contain $\frac{\mathbf{v}}{c}$. 
To remove speed of light by redefinition of units is unnatural.
\item The Lorentz force  is a relativistic effect,
so the formula \emph{have to} contain $\frac{\mathbf{v}}{c}$.
To remove speed of light by redefinition of units is unnatural.
\item Introducing of constant $\frac1{4\pi}$ into the Coulomb law
(it was suggested by O. Heaviside) is natural, because it removes
the factor $4\pi$ (the surface area of two-dimensional unit sphere)
in Maxwell equations and in formulae for energy and action of
electromagnetic field.
It is consistent with the practice of theorists which consider spaces with dimensions other then 3.
For the standard electrodynamics this rationalisation makes no bad nor good.
\end{itemize}

\subsection{The wishes of an engineer and an experimenter}

\begin{itemize}
\item The SI units for electrical circuits (ampere, volt, ohm, farad, henry) are used everywhere in devices and standards, they could not be changed.
\item The appearance of the speed of light in equations for electrical circuits is undesirable.
\item Fields $\mathbf{D}$ and $\mathbf{H}$ could not be measured directly, so 
their units are not used in any devices. This units are not very important.
\end{itemize}

\subsection{How to reconcile theorists with engineers}

We propose to modify the SI without changing the base units 
(kilogram, metre, second, ampere, mole, kelvin, candela), 
but modifying the form of the equations of electrodynamics 
(to make them similar to equations in the CGS)
by changing the constant factors.
Due to this change of factors some SI derived units have to be changed.

We preserve from old SI ($\text{SI}_\text{old}$) ampere as unit for current,
and all derived units, which do not involve fields $\mathbf{D}$, $\mathbf{B}$ and $\mathbf{H}$.
So the units of charge $\text{coulomb}=\text{C}$, 
of electric potential $\text{volt}=\text{V}$,
of electrical resistance $\text{ohm}=\Omega$,
of capacitance $\text{farad}=\text{F}$,
of inductance $\text{henri}=\text{H}$
remain the same.

The Coulomb law and the strength of electrical field reman the same form as in $\text{SI}_\text{old}$
$$
  F=\frac1{4\pi\varepsilon_0}\frac{q_1q_2}{r^2}=k_e\frac{q_1q_2}{r^2},\qquad
  E=\frac1{4\pi\varepsilon_0}\frac{q}{r^2}=k_e\frac{q}{r^2}.
$$
Here $k_e=\frac1{4\pi\varepsilon_0}$ is \emph{Coulomb constant}, $q_1,q_2,q$ are electrical charges.

We redefine the fields $\mathbf{D}$, $\mathbf{H}$ and $\mathbf{B}$
according to above ``The wishes of a theorist'', all these fields
have the same unit $\frac{\text{V}}{\text{m}}$, just like field $\mathbf{E}$.

\textbf{All formulae and units for electrical circuits (without fields)
preserve the same form as in the $\textbf{SI}_\textbf{old}$.}

\textbf{All formulae (but not units!) for electromagnetic fields look similar to
formulae of the CGS.}

This future version of the SI we call Physics and Technology System of units (PT).
The name of the PT system based on the idea to combine the advantages of systems used in physics and engineering.

During the process of preparing the paper it was discovered that this system of units has already been proposed previously
(from 2001) by G.M.~Trunov \cite{trunov2011,trunov1,trunov2,trunov3}.
He calls it \emph{theoretical system of electromagnetic units} (SI(T)).
Unfortunately this initiative has not yet received wide distribution.
We think, to change the name of system of units in accordance with the terminology of G.M.~Trunov is not feasible, 
because for the expansion of the new system its benefits in the convergence of physics 
and engineering are more important than theoretical advantages (the theorists are quite happy with the CGS).

\section{The physics and technology system of units (PT)}
\subsection{Conversion from the CGS to the PT}
The base units of the PT system coincide with the base units of $\text{SI}_\text{old}$.
So, using the standard ($\text{SI}_\text{old}$) equations for
current $I$, electrical potential $U$, power $\frac{d\mathcal{E}}{dt}$,
energy $\mathcal{E}_C$ and $\mathcal{E}_L$ for capacitance $C$ and inductance $L$
$$
  I=\frac{dq}{dt},\qquad
  UI=\frac{d\mathcal{E}}{dt},\qquad
  R=\frac{U}{I},\qquad
  \mathcal{E}_C=\frac{q^2}{2C},\qquad
  \mathcal{E}_L=\frac{LI^2}2,
$$
 one derives the units of $q$, $U$, $C$, $L$ in terms of kilogram, metre, second and ampere.
 All these units coincide with units of $\text{SI}_\text{old}$.\footnote{The 
 only variable related with field here is $U$.
 All variables defined in terms of circuit parameters.
 According to elecromechanical analogy charge $q$ is generalised coordinate,
 current $I$ is generalised velocity,
 potential difference $U$ is generalised force,
 resistance $R$ is viscous friction constant,
 inverse capacitance $\frac1C$ is spring constant,
 inductance $L$ is mass constant.
 Similarly to $\text{SI}_\text{old}$ speed of light does not involved in
 electrical circuit equations.}

  We derive the electromagnetic field equations in the PT system by modification of
  the CGS equations.

 In Gaussian CGS the particular choice of centimetre gramme and second as base units
 is not significant. We are interested in just the form of equations (the choice of constants). 
 So, let us introduce an auxiliary Gaussian system of units MKS (metre, kilogram, second Gaussian, MKSG).
 All equations of MKSG has the same form as the equations of Gaussian CGS.

  From the point of view of MKSG the Coulomb constant 
  (hereinafter, the braces denote numerical value of dimensional quantity)
$$
  k_e=\frac1{4\pi\varepsilon_0},\qquad\{k_e\}=\{c\}^2\times 10^{-7},
$$
 is used just to convert charge units from SI to MKSG
$$
  F=\frac1{4\pi\varepsilon_0}\frac{q_{1si}q_{2si}}{r^2}=k_e\frac{q_{1si}q_{2si}}{r^2}
  =\frac{q_{1mksg}\,q_{2mksg}}{r^2},\qquad
  q_{mksg}=\sqrt{k_e}\,q_{si}.
$$

  If the force law for a charge $q$ in electric field $\mathbf{E}$
 does not contain any coefficients, then
$$
  \mathbf{F}=q_{mksg}\mathbf{E}_{mksg}=q_{si}\mathbf{E}_{si},\qquad
  \mathbf{E}_{mksg}=\frac{\mathbf{E}_{si}}{\sqrt{k_e}}.
$$
 For electrical charges, currents and field $\mathbf{E}$
 there is no difference between the PT system and the $\text{SI}_\text{old}$.

  In the MKSG, like in the CGS all fields $\mathbf{E},\mathbf{D},\mathbf{B},\mathbf{H}$
  have the same units. To preserve this property in PT system we use 
  the same conversion factor to convert all four fields from the MKSG to the PT.
  Also we use the equal conversion factors to convert all field sources
  (charges, currents, electric and magnetic multipoles).

  I.e. we make the following substitution into the electrodynamics equations
  of the MKSG (which look like exactly as equations of the CGS)
$$
  (q,\mathbf{j},\dots)_{mksg}=\sqrt{k_e}\times(q,\mathbf{j},\dots)_{pt},
$$
$$
  (\mathbf{E},\mathbf{D},\mathbf{B},\mathbf{H})_{mksg}=\frac1{\sqrt{k_e}}
  \times(\mathbf{E},\mathbf{D},\mathbf{B},\mathbf{H})_{pt}.
$$
So we have only two different \emph{conversion factor}
from the PT system to the usual for theorists Gaussian system (not CGS, but MKSG):
one conversion factor for all fields ($\mathbf{E},\mathbf{D},\mathbf{B},\mathbf{H}$), 
the other for all sources (charges, currents, multipoles).
The product of the conversion factors is 1\footnote{The substitution
have to be done only in the fields equations. The form of the equations of circuits and the definitions
of multipoles is the same as in Gaussian system.}.

You don't need a whole table of conversion factors, you use to convert between the $\text{SI}_\text{old}$ and the CGS!

\subsection{Equations in the PT system}
The electrodynamics equations in the PT system have the following form
\bean
\Div\mathbf{B}&=&0,\quad
\rot\mathbf{E}=-\frac1c\frac{\p\mathbf{B}}{\p t},\\
\Div\mathbf{D}&=&4\pi k_e\rho,\quad
\rot\mathbf{H}=\frac1c\frac{\p\mathbf{D}}{\p t}+\frac{4\pi}{c}k_e\mathbf{j},\\
\mathbf{S}&=&\frac{c}{4\pi k_e}[\mathbf{E}\times\mathbf{H}],\quad
W=\frac{(\mathbf{E},\mathbf{D})+(\mathbf{B},\mathbf{H})}{8\pi k_e},\\
\sigma_{\alpha\beta}&=&\frac{E_\alpha D_\beta+B_\alpha H_\beta}{4\pi k_e}-\delta_{\alpha\beta}W,\\
\mathbf{F}&=&q\left(\mathbf{E}+\frac1c[\mathbf{v}\times\mathbf{B}]\right),\qquad
L=\frac{\Phi}{cI},\\
\mathbf{D}&=&\mathbf{E}+4\pi k_e \mathbf{P},\qquad
\mathbf{H}=\mathbf{B}-4\pi k_e\mathbf{M}.
\eean
Here $\rho$ and $\mathbf{j}$ are densities of charge and current,
$W$ is density of field energy,
$\mathbf{S}$ is the Umov-Pointing vector,
$\sigma_{\alpha\beta}$ is the Maxwell stress tensor,
indices $\alpha,\beta=1,2,3$ numerate space coordinates,
$\Phi$ is the magnetic flux,
$\mathbf{P}$ and $\mathbf{M}$ are electric and magnetic dipoles per unit volume.

The substitution $k_e\to1$ reproduce the equations in the Gaussian form.
The only exception is inductance, which sometimes defined in the CGS as $L_{cgs}=\frac{\Phi}{I}$.
Nevertheless this is not shortcoming, but the advantage, because it removes speed of light
from the electrical circuit equations.

The conversion from CGS formulae to PT ones (the insertion of the Coulomb constant)
is easily performed using dimensionalities.
This conversion is not much more difficult, than the insertion of the Boltzmann constant, which was taken to be 1.

The fields in the $\text{SI}_\text{old}$ and in the PT system are converted by the following relations
$$
\mathbf{E}_{pt}=\mathbf{E}_{si_{old}},\qquad 
\mathbf{D}_{pt}=\frac{\mathbf{D}_{si_{old}}}{\varepsilon_0}=4\pi k_e \mathbf{D}_{si_{old}},
$$
$$
\mathbf{B}_{pt}=c\mathbf{B}_{si_{old}},\qquad 
\mathbf{H}_{pt}=\frac{\mathbf{H}_{si_{old}}}{c\varepsilon_0}=c\mu_0\mathbf{H}_{si_{old}}
=\frac{4\pi}{c}k_e\mathbf{H}_{si_{old}},
$$
$$
\varepsilon_0=\frac{1}{4\pi k_e},\qquad
\mu_0=\frac1{c^2\varepsilon_0}=\frac{4\pi k_e}{c^2}.
$$

To comply with the old system $\text{SI}_\text{old}$ it is useful to
introduce an \emph{``effective'' magnetic field}, which coincides with
the magnetic field of $\text{SI}_\text{old}$
$$
  \mathbf{B}_{eff}=\frac{\mathbf{B}_{pt}}{c}=\mathbf{B}_{si_{old}}
$$
The modern ($\text{SI}_\text{old}$) devices are calibrated in the
units of effective magnetic field, however, the use of it in formulae is 
undesirable.\footnote{It is easy to remember, if one remembers the fact that the Lorentz force
written in terms of effective magnetic field does not involve the speed of light.}.

\section{Conclusion}

In 2011 The General Conference on Weights and Measures (CGPM), at its 24th meeting,
adopted the Resolution ``On the possible future revision of the International System of Units, the SI''
\cite{resolution2011}. 
The resolution recommends to change the definition of kilogram and ampere
by simultaneous fixing the exact numerical values of the Planck constant and of the elementary charge.

It means that the base SI units will be defined by quantum electrodynamics effects.
The 8-th edition of ``SI brochure'' \cite{si8} has already recognized the advantages of 
``the use of CGS-Gaussian units in electromagnetic theory applied to
quantum electrodynamics and relativity'', so the simultaneous introducing of
the same unit $\frac{\text{V}}{\text{m}}$ for all four fields 
$\mathbf{E}$, $\mathbf{D}$, $\mathbf{B}$, $\mathbf{H}$
(i.e. the introducing of the PT system as new version of the SI) looks natural and feasible.

This reform of the SI would be invisible for the general public, because
ampere, volt, ohm, farad and henry remain the same.
For the majority of physicists and engineers this reform would not be a big problem, because
electrical circuits units remains the same as in the SI, and new field equations are close to the CGS equations.

The century dispute on the choice of units for electrodynamics could be resolved in next few years, 
if the community of physicists to show an active position.

\section*{Acknowledgements}

The author thanks Yu.R.~Alanakyan, A.L.~Barabanov, V.S.~Bulygin, V.V.~Ezhela, A.D.~Gladun,
A.N~Kirichenko, S.M.~Kozel, A.S.~Leskov, V.A.~Ovchinkin, A.A.~Pukhov, Ya.A.~Romanyuk, A.A.~Rukhadze, 
V.P.~Slobodyanin, V.P.~Tarakanov, G.M.~Trunov, S.V.~Vinogradov, V.G.~Zhotikov 
and other faculty of the Department of Theoretical Physics
and the Department of General Physics of Moscow Institute of Physics and Technology, and participants
of the seminar of the Theoretical Department of Prokhorov General Physics Institute of
Russian Academy of Sciences for the useful discussions and valuable criticism.

\appendix
\renewcommand{\thesection}{\Alph{section}.}

\section{The SI in electrodynamics}
Let us present for comparison the same key equations of electrodynamics in the SI units:
\bean
\Div\mathbf{B}&=&0,\quad
\rot\mathbf{E}=-\frac{\p\mathbf{B}}{\p t},\\
\Div\mathbf{D}&=&\rho,\quad
\rot\mathbf{H}=\frac{\p\mathbf{D}}{\p t}+\mathbf{j},\\
\mathbf{S}&=&[\mathbf{E}\times\mathbf{H}],\quad
W=\frac1{2}(\mathbf{E},\mathbf{D})+\frac1{2}(\mathbf{B},\mathbf{H}),\\
\sigma_{\alpha\beta}&=&\left(E_\alpha D_\beta+B_\alpha H_\beta\right)-\delta_{\alpha\beta}W,\\
\mathbf{F}&=&q\left(\mathbf{E}+[\mathbf{v}\times\mathbf{B}]\right),\qquad
L=\frac{\Phi}{I},\\
\mathbf{D}&=&\varepsilon_0\mathbf{E}+\mathbf{P},\qquad
\mathbf{H}=\frac{\mathbf{B}}{\mu_0}-\mathbf{M}.
\eean
One could think, that the equations in the SI look simpler.
To dispel this illusion, and to explain the reason for the preference of theorists, 
we write the SI equation in vacuum (fields $\mathbf{D}$ and $\mathbf{H}$ are excluded)
\bean
\Div\mathbf{B}&=&0,\quad
\rot\mathbf{E}=-\frac{\p\mathbf{B}}{\p t},\\
\Div\mathbf{E}&=&\frac1{\varepsilon_0}\rho,\quad
\rot\mathbf{B}=\varepsilon_0\mu_0\frac{\p\mathbf{E}}{\p t}+\mu_0\mathbf{j}
=\frac1{c^2}\frac{\p\mathbf{E}}{\p t}+\frac1{\varepsilon_0c^2}\mathbf{j},\\
\mathbf{S}&=&\frac1{\mu_0}[\mathbf{E}\times\mathbf{B}]
=\varepsilon_0c^2[\mathbf{E}\times\mathbf{B}],\quad
W=\frac{\varepsilon_0}{2}\mathbf{E}^2+\frac1{2\mu_0}\mathbf{B}^2
=\frac{\varepsilon_0}{2}(\mathbf{E}^2+c^2\mathbf{B}^2),\\
\sigma_{\alpha\beta}&=&\left(\varepsilon_0E_\alpha E_\beta
+\frac1{\mu_0}B_\alpha B_\beta\right)-\delta_{\alpha\beta}W
=\varepsilon_0\left(E_\alpha E_\beta
+c^2B_\alpha B_\beta\right)-\delta_{\alpha\beta}W.
\eean

\section{The equations of PT system with Heaviside constant}
 Some physicists prefer the Heaviside-Lorentz units.
 Let us present the PT system electrodynamics equations with 
 \emph{Heaviside constant}
$$
  \varkappa_e=\frac1{\varepsilon_0}=4\pi k_e,\qquad
  \{\varkappa_e\}=4\pi \{c\}^2\times 10^{-7}.
$$
The same key equations of electrodynamics in the PT system with Heaviside constant have the following form
\bean
\Div\mathbf{B}&=&0,\quad
\rot\mathbf{E}=-\frac1c\frac{\p\mathbf{B}}{\p t},\\
\Div\mathbf{D}&=&\varkappa_e\rho,\quad
\rot\mathbf{H}=\frac1c\frac{\p\mathbf{D}}{\p t}+\frac{\varkappa_e}{c}\mathbf{j},\\
\mathbf{S}&=&\frac{c}{\varkappa_e}[\mathbf{E}\times\mathbf{H}],\quad
W=\frac{(\mathbf{E},\mathbf{D})+(\mathbf{B},\mathbf{H})}{2\varkappa_e},\\
\sigma_{\alpha\beta}&=&\frac1{\varkappa_e}\left(E_\alpha D_\beta+B_\alpha H_\beta\right)-\delta_{\alpha\beta}W,\\
\mathbf{F}&=&q\left(\mathbf{E}+\frac1c[\mathbf{v}\times\mathbf{B}]\right),\qquad
L=\frac{\Phi}{cI},\\
\mathbf{D}&=&\mathbf{E}+\varkappa_e\mathbf{P},\qquad
\mathbf{H}=\mathbf{B}-\varkappa_e\mathbf{M}.
\eean

The substitution $\varkappa_e\to1$ reproduce the equations in the Heaviside-Lorentz form
(in the MKS-Heavisidean system, MKSH).
The inverse conversion is
$$
  (q,\mathbf{j},\dots)_{mksh}=\sqrt{\varkappa_e}\times(q,\mathbf{j},\dots)_{pt},
$$
$$
  (\mathbf{E},\mathbf{D},\mathbf{B},\mathbf{H})_{mksh}=
  \frac1{\sqrt{\varkappa_e}}
  \times(\mathbf{E},\mathbf{D},\mathbf{B},\mathbf{H})_{pt}.
$$

\section{The meaning of the Coulomb constant}

The tradition of the Gaussian system of units assumes that 
the Coulomb constant has no physical meaning, 
since it represents just the square of the ratio of two units of charge:
$$
  \{k_e\}=\left(\frac{q_{mksg}}{1~\text{C}}\right)^2,
$$
here
$$
  q_{mksg}=\sqrt{\text{J}\cdot\text{m}}=\text{kg}^{1/2}\cdot\text{m}^{3/2}\cdot\text{s}^{-1}=
  10^{4.5}~\text{CGS unit of charge}
$$
is the MKSG unit of charge.

In the modern SI, in which ampere (and coulomb as a derived unit) is defined via the 
force acting between two parallel conductors with a current, 
the opinion of $k_e$ physical meaningless is absolutely true.

In 2011 The General Conference on Weights and Measures (CGPM), at its 24th meeting,
adopted the Resolution ``On the possible future revision of the International System of Units, the SI''
\cite{resolution2011}. 
The resolution recommends to redefine kilogram by fixing the exact numerical value of the Planck constant
(similarly metre was redefined by fixing the exact numerical value of the speed of light)
$$
h=2\pi\hbar=6.626\,06\mathrm{X}\times10^{-34}~\text{J}\cdot\text{s},
$$
and to redefine ampere by fixing the exact numerical value of the elementary charge\footnote{It would be natural
to require that 1~C were an integer number of $e$.}
$$
e=1.602\,17\mathrm{X}\times10^{-19}~\text{C}.
$$
In both cases $\mathrm{X}$ denotes some digits, which have to be fixed in future.

In year 1983 the metre was redefined by fixing the exact numerical value of the speed of light.
Now \emph{by definition} speed of light has exact numerical value:
$$
 c=299\,792\,458~\frac{\text{m}}{\text{s}}.
$$

So, in the definition of the fine-structure constant\footnote{The fine-structure constant
itself is the square of the ratio of two \emph{natural} units of charge, the elementary charge
and the Plank charge
$\alpha=\left(\frac{e^2}{\hbar
c}\right)_{cgs}=\left(\frac{e}{q_{Pl}}\right)^2\approx\frac1{137{,}035\,999}$.}
$$
  \alpha=\frac{e^2}{4\pi\varepsilon_0\hbar c}=\frac{k_e e^2}{\hbar c}
$$
all quantities will be defined \emph{by exact numerical values by definition},
except of one, the Coulomb constant.
$$
 k_e=\frac{\hbar c}{e^2}\cdot\alpha=C_\alpha\cdot\alpha.
$$

We see that the numerical value of $k_e$ in the future reformed SI 
is the fine structure constant, multiplied by some fixed numerical factor. 
The factor is chosen so that the redefined units of mass and current reproduced 
taken thus far with the accuracy available with modern 
(at the time of the reform) measuring devices.

The further improvement of the accuracy of the measurements will 
inevitably makes the difference of units of the modern 
SI and the reformed SI available for measurement. 
With this moment the Coulomb constant becomes really fundamental constant.
It is the fine structure constant multiplied by historically
fixed factor $C_\alpha$.

Thus, the Coulomb constant acquires a physical meaning due to the fact 
that we have a natural charge unit (the elementary charge), which
is not a multiple of Planck charge.

The dispute between the Gaussian CGS system (and the old SI) and the future SI on definition 
of charge unit is related to the following simple choice.
Do we prefer to define charge unit using the Plank charge (the CGS system and the old SI),
or using the elementary charge (the new SI).
Since the Planck charge in nature is not implemented, the choice seems to be obvious.

\section{The practical units-1990 and the redefinition of the ampere and the kilogram}
This appendix is based mainly on materials of the paper by S.G.~Karshenboym
``On the redefinition of the kilogram and ampere in terms of fundamental physical constants''
\cite{karshenboim}.
The possible modernizations of the SI are discussed in paper \cite{modernizing}.

\emph{Simultaneous} redefinition of kilogram and ampere by fixing exact numerical values of the Plank constant
and the elementary charge is related with the following fact.
The modern official definition of ampere using the force of interaction between two 
parallel currents.
It corresponds to fixing of exact numerical value of
$$
\mu_0=\frac{\varepsilon_0}{c^2}=4\pi\times10^{-7}~\frac{\text{H}}{\text{m}}.
$$
This official definition is less precise than definition of ampere using quantum effects.
Quantum effects are used in definitions of ``practical'' units of year 1990, the volt-1990 and the ohm-1990.

The volt-1990 is defined using the Josephson effect.
During the flow of superconducting current through Josephson junction the extra energy $2eU$
received by a Cooper pair from external potential difference is emitted in the form of photons.
The photon frequency is $\nu=\frac{2e}{h}\,U$.
The definition of the volt-1990 fixes exact numerical value of the Josephson constant
$$
 K_J=\frac{2e}h=483\,597.9~\frac{\text{GHz}}{\text{V}_{90}}.
$$

The ohm-1990 is defined using the integer quantum Hall effect, which is observed in 
2-dimensional degenerate electron gas at low temperature in magnetic field perpendicular to 2d gas plane.
Across the direction of current, a voltage proportional to the current appears $U_y=\rho_{xy}I_x$.
The Hall resistance is $\rho_{xy}=\frac{R_K}{n}$, where $n$ is an integer number.
The definition of the ohm-1990 fixes the exact numerical value of the von Klitzing constant
$$
 R_K=\frac{h}{e^2}=25\,812.807~\Omega_{90}.
$$

So, the \emph{simultaneous} fixing of the Josephson constant and the von Klitzing constant
defines the exact numerical values of the Plank constant and of the elementary charge.
The units of volt-1990, ohm-1990, metre and second are the base units of the \emph{practical system}.
The practical system is an alternative version of SI.
It has higher precision, but it is different from the SI.
The ``practical kilogram'' differ from the standard kilogram by $10^{-7}$, the ``practical constant $\mu_0$''
is not a fixed value.

For some modern measurement the main source of inaccuracy is not measurement itself, but the definition of measurement unit.
Practical units are defined more precisely than the SI units, so the redefinition of kilogram and ampere according
to Resolution of year 2011 looks useful.
It does not mean introducing of volt-1990 and ohm-1990 as the new SI standards, because since the year 1990 
measurement technic has improved.
The discrepancy of ``practical kilogram'' with the standard kilogram by $10^{-7}$ is no longer acceptable.

\section{The notion of the system of the systems of units}

Some systems of units (the SI, the system of practical units) are the systems preferably for measurements 
and the others (the CGS, the Heaviside-Lorentz system and some other systems) are the ``theoretical'' systems, preferably
for formulae and analytical calculations.
The units of theoretical systems are sometimes inconvenient for actual measurements.

The convenient method to introduce a theoretical system is to set some
constants to be 1. E.g. theorists prefer to set the Boltzmann constant to be 1.
It is convenient for a theorist, but it is unusable for an experimenter.
Nevertheless, after the derivation of final formula, a theorist could
easily reintroduce the Boltzmann constant using the dimensionalities of physical
quantities.
Similarly in the special relativity system of units the speed of light is set to be 1,
in the general relativity system the gravitational constant too.
In the quantum mechanical system the Plank constant $\hbar$ is set to be 1,
in the atomic physics system the elementary charge and the electron mass too.
In all these cases the inverse conversion to experimental units
could be done easily by using the dimensionalities of physical quantities.

We have the natural \emph{system of the systems} of theoretical units, which
originates from Gaussian CGS\footnote{The other system of the systems of theoretical units 
originates from the Heaviside-Lorentz system of units.}. 
The other theoretical systems (convenient for
different theoretical physics branches) generates from
CGS by substituting 1 instead of different constants.
And so on until completely dimensionless system of units like the Planck units.

Unfortunately the form of electrodynamics equations in the CGS and the $\text{SI}_\text{old}$ is different.
So the $\text{SI}_\text{old}$ is not connected with the system of theoretical systems.
The rules of conversion between the $\text{SI}_\text{old}$ and the CGS are to complicated.

The PT system naturally becomes in the root of the system of theoretical systems.
To convert from the PT system to the Gaussian system one just set the Coulomb constant to be 1.
It connects the systems of theoretical systems with experiment.\footnote{The other 
system of the theoretical systems originates from the Heaviside-Lorentz system is also
naturally connected with the PT. 
To convert from the PT system to the Heavisidian system one just set the Heaviside constant to be 1.}

\section{RPP-2014 electromagnetic relations}
\emph{To demonstrate the convenience of the PT system,
we reproduce here the section 7 ELECTROMAGNETIC RELATIONS of The Review of Particle Physics \cite{rpp2014}, using PT units.
We expand the table below to add the PT system.
Two subsections, which are exactly the same in PT units and in SI units are presented by their headlines and short comments only.
The third subsection is rewritten from CSG system to PT system 
(actually we just insert $k_e$ constant in two formulae of the 3rd subsection).}

\section*{7. Electromagnetic relations}
~\\
{\scriptsize
\begin{tabular}{|l|l|l|l|}
\hline\hline&&&\\
Quantity&Gaussian CGS&SI&PT system\\&&&\\
\hline&&&\\
Conversion factors:&&&\\&&&\\
~~Charge:&$2.997\,924\,58\times10^9$~esu& = 1~C = 1~A\,s& = 1~C = 1~A\,s\\&&&\\
~~Potential:&(1/299.792\,458) statvolt  (ergs/esu)& = 1~V = 1~J\,C${}^{-1}$& = 1~V = 1~J\,C${}^{-1}$\\&&&\\
~~Magnetic field:&$10^4$~gauss = $10^4$~dyne/esu& = 1~T = 1~N A${}^{-1}$m${}^{-1}$& = $2.997\,924\,58\times10^8$~V/m\\&&&\\
\hline&&&\\
&$\mathbf{F}=q(\mathbf{E}+\frac{\mathbf{v}}{c}\times\mathbf{B})$&$\mathbf{F}=q(\mathbf{E}+\mathbf{v}\times\mathbf{B})$&
$\mathbf{F}=q(\mathbf{E}+\frac{\mathbf{v}}{c}\times\mathbf{B})$\\&&&\\
\hline&&&\\
&$\boldsymbol\nabla\bullet\mathbf{D}=4\pi\rho$&
$\boldsymbol\nabla\bullet\mathbf{D}=\rho$&
$\boldsymbol\nabla\bullet\mathbf{D}=4\pi k_e\rho$\\&&&\\
&$\boldsymbol\nabla\times\mathbf{H}-\frac1c\frac{\p\mathbf{D}}{\p t}=\frac{4\pi}{c}\mathbf{J}$
&$\boldsymbol\nabla\times\mathbf{H}-\frac{\p\mathbf{D}}{\p t}=\mathbf{J}$
&$\boldsymbol\nabla\times\mathbf{H}-\frac1c\frac{\p\mathbf{D}}{\p t}=\frac{4\pi}{c}k_e\mathbf{J}$\\&&&\\
&$\boldsymbol\nabla\bullet\mathbf{B}=0$
&$\boldsymbol\nabla\bullet\mathbf{B}=0$
&$\boldsymbol\nabla\bullet\mathbf{B}=0$\\&&&\\
&$\boldsymbol\nabla\times\mathbf{E}-\frac1c\frac{\p\mathbf{H}}{\p t}=0$
&$\boldsymbol\nabla\times\mathbf{E}-\frac{\p\mathbf{H}}{\p t}=0$
&$\boldsymbol\nabla\times\mathbf{E}-\frac1c\frac{\p\mathbf{H}}{\p t}=0$\\&&&\\
\hline&&&\\
Constitutive relations:&$\mathbf{D}=\mathbf{E}+4\pi\mathbf{P}$, 
&$\mathbf{D}=\varepsilon_0\mathbf{E}+\mathbf{P}$, 
&$\mathbf{D}=\mathbf{E}+4\pi k_e\mathbf{P}$, \\&&&\\
&$\mathbf{H}=\mathbf{B}-4\pi\mathbf{M}$
&$\mathbf{H}=\mathbf{B}/\mu_0-\mathbf{M}$
&$\mathbf{H}=\mathbf{B}-4\pi k_e\mathbf{M}$\\&&&\\
\hline&&&\\
Linear media:&$\mathbf{D}=\epsilon\mathbf{E}$, $\mathbf{H}=\mathbf{B}/\mu$
&$\mathbf{D}=\epsilon\mathbf{E}$, $\mathbf{H}=\mathbf{B}/\mu$
&$\mathbf{D}=\epsilon\mathbf{E}$, $\mathbf{H}=\mathbf{B}/\mu$\\&&&\\
&1&$\epsilon_0=8.854\,187\dots\times10^{-12}~\text{F}\,\text{m}^{-1}$&1\\&&&\\
&1&$\mu_0=4\pi\times10^{-7}~\text{N}\,\text{A}^{-2}$&1\\&&&\\
\hline&&&\\
&$\mathbf{E}=-\boldsymbol\nabla V-\frac1c\frac{\p\mathbf{A}}{\p t}$
&$\mathbf{E}=-\boldsymbol\nabla V-\frac{\p\mathbf{A}}{\p t}$
&$\mathbf{E}=-\boldsymbol\nabla V-\frac1c\frac{\p\mathbf{A}}{\p t}$\\&&&\\
&$\mathbf{B}=\boldsymbol\nabla\times\mathbf{A}$
&$\mathbf{B}=\boldsymbol\nabla\times\mathbf{A}$
&$\mathbf{B}=\boldsymbol\nabla\times\mathbf{A}$\\&&&\\
\hline&&&\\
&$V=\sum\limits_{charges}\frac{q_i}{r_i}=\int\frac{\rho(\mathbf{r'})\,d^3x'}{|\mathbf{r}-\mathbf{r'}|}$
&$V=\frac1{4\pi\epsilon_0}\sum\limits_{ch.}\frac{q_i}{r_i}=\frac1{4\pi\epsilon_0}\int\frac{\rho(\mathbf{r'})\,d^3x'}{|\mathbf{r}-\mathbf{r'}|}$
&$V=k_e\sum\limits_{ch.}\frac{q_i}{r_i}=k_e\int\frac{\rho(\mathbf{r'})\,d^3x'}{|\mathbf{r}-\mathbf{r'}|}$\\&&&\\
&$\mathbf{A}=\frac1c\oint\frac{Id\boldsymbol\ell}{|\mathbf{r}-\mathbf{r'}|}=\frac1c\int\frac{\mathbf{J}\,d^3x'}{|\mathbf{r}-\mathbf{r'}|}$
&$\mathbf{A}=\frac{\mu_0}{4\pi}\oint\frac{Id\boldsymbol\ell}{|\mathbf{r}-\mathbf{r'}|}=\frac{\mu_0}{4\pi}\int\frac{\mathbf{J}\,d^3x'}{|\mathbf{r}-\mathbf{r'}|}$
&$\mathbf{A}=\frac{k_e}c\oint\frac{Id\boldsymbol\ell}{|\mathbf{r}-\mathbf{r'}|}=\frac{k_e}c\int\frac{\mathbf{J}\,d^3x'}{|\mathbf{r}-\mathbf{r'}|}$\\
&&&\\
\hline&&&\\
&$\mathbf{E}'_\|=\mathbf{E}_\|$&$\mathbf{E}'_\|=\mathbf{E}_\|$&$\mathbf{E}'_\|=\mathbf{E}_\|$\\&&&\\
&$\mathbf{E}'_\bot=\gamma(\mathbf{E}_\bot+\frac1c\mathbf{v}\times\mathbf{B})$
&$\mathbf{E}'_\bot=\gamma(\mathbf{E}_\bot+\mathbf{v}\times\mathbf{B})$
&$\mathbf{E}'_\bot=\gamma(\mathbf{E}_\bot+\frac1c\mathbf{v}\times\mathbf{B})$\\&&&\\
&$\mathbf{B}'_\|=\mathbf{B}_\|$&$\mathbf{B}'_\|=\mathbf{B}_\|$&$\mathbf{B}'_\|=\mathbf{B}_\|$\\&&&\\
&$\mathbf{B}'_\bot=\gamma(\mathbf{B}_\bot-\frac1c\mathbf{v}\times\mathbf{E})$
&$\mathbf{B}'_\bot=\gamma(\mathbf{B}_\bot-\frac1{c^2}\mathbf{v}\times\mathbf{E})$
&$\mathbf{B}'_\bot=\gamma(\mathbf{B}_\bot-\frac1c\mathbf{v}\times\mathbf{E})$\\&&&\\
\hline
\multicolumn{4}{|c|}{~}\\
\multicolumn{4}{|c|}{$k_e=\frac{1}{4\pi\epsilon_0} =c^2\times10^{-7}~\text{N}\,\text{A}^{-2}=8.987\,55\dots\times10^9~\text{m}\,
\text{F}^{-1};~~~~
\frac{\mu_0}{4\pi}=10^{-7}~\text{N}\,\text{A}^{-2};~~~~
c=\frac1{\sqrt{\epsilon_0\mu_0}}=2.997\,924\,58\times10^8~\text{m}\,\text{s}^{-1}$}\\
\multicolumn{4}{|c|}{~}\\
\hline\hline
\end{tabular}}

\subsection*{7.1. Impedances (SI units \emph{= PT inits})}
\emph{The original text uses SI. The formulae in PT are exactly the same.}

\subsection*{7.2. Capacitors, inductors, and transmission Lines (SI units \emph{= PT inits})}
\emph{The original text uses SI. The formulae in PT are exactly the same.}

\subsection*{7.3. Synchrotron radiation (CGS units \emph{and PT units})}
\emph{We use the PT system in this subsection.
To convert formulae from PT back to CGS just set $k_e=1$.}

For a particle of charge $e$, velocity $v=\beta c$, and energy $E =\gamma mc^2$,
traveling in a circular orbit of radius $R$, the classical energy loss per
revolution $\delta E$ is
\be
  \delta E=\frac{4\pi}{3}\frac{k_e e^2}{r}\beta^3\gamma^4.
\ee
For high-energy electrons or positrons ($\beta\approx 1$), this becomes
\be
\delta E \text{(in MeV)}\approx 0.0885 [E\text{(in GeV)}]^4/R\text{(in m)}.
\ee
For $\gamma\gg 1$, the energy radiated per revolution into the photon energy
interval $d(\hbar\omega)$ is
\be\label{rpp-dI}
  dI = \frac{8\pi}9 \alpha\gamma\, F(\omega/\omega_c)\, d(\hbar\omega),
\ee
where $\alpha = k_ee^2/\hbar c$ is the fine-structure constant and
\be
  \omega_c=\frac{3\gamma^3c}{2R}
\ee
is the critical frequency. The normalized function $F(y)$ is
\be
  F(y)=\frac{9}{8\pi}\sqrt3 y\int_y^\infty K_{5/3}(x)\,dx,
\ee
where $K_{5/3}(x)$ is a modified Bessel function of the third kind. For
electrons or positrons,
\be
\hbar\omega_c \text{(in keV)}\approx 2.22 [E\text{(in GeV)}]^3/R\text{(in m)}.
\ee
For $\gamma\gg1$ and $\omega\ll\omega_c$,
\be
  \frac{dI}{d(\hbar\omega)}\approx 3.3\alpha (\omega R/c)^{1/3},
\ee
whereas for
$$
  \gamma\gg1~~~\text{and} ~~~\omega\gtrsim 3\omega_c,
$$
\be
  \frac{dI}{d(\hbar\omega)}\approx\sqrt{\frac{3\pi}{2}}\alpha\gamma\left(\frac{\omega}{\omega_c}\right)^{1/2}
  e^{-\omega/\omega_c}\left[1+\frac{55}{72}\frac{\omega}{\omega_c}+\cdots\right].
\ee
The radiation is confined to angles $\lesssim1/\gamma$ relative to the instantaneous
direction of motion. For $\gamma\gg1$, where Eq. \eqref{rpp-dI} applies, the mean
number of photons emitted per revolution is
\be
  N_\gamma=\frac{5\pi}{\sqrt3}\alpha\gamma,
\ee
and the mean energy per photon is
\be
\langle\hbar\omega\rangle=\frac{8}{15\sqrt3}\hbar\omega_c.
\ee
When $\langle\hbar\omega\rangle\gtrsim\text{O}(E)$, quantum corrections are important.

\end{document}